\documentclass[twocolumn,superscriptaddress,amsmath,amssymb,aps,prl,reprint,floatfix]{revtex4-2}

\usepackage{graphicx}
\usepackage{dcolumn}
\usepackage{bm}
\usepackage{float}

\begin{document}

\title{Mapping the Electronic Structure of Warm Dense Nickel via Resonant Inelastic X-ray Scattering}

\author{O.S.~\surname{Humphries}}
\affiliation{Department of Physics, Clarendon Laboratory, University of Oxford, Parks Road, Oxford, OX1 3PU, UK}

\author{R.S.~\surname{Marjoribanks}}
\affiliation{Department of Physics, University of Toronto, 60 St. George Street, Toronto, Ontario, M5S 1A7, Canada}

\author{Q.Y.~\surname{van den Berg}}
\affiliation{Department of Physics, Clarendon Laboratory, University of Oxford, Parks Road, Oxford, OX1 3PU, UK}

\author{E.C.~\surname{Galtier}}
\affiliation{Linac Coherent Light Source, SLAC National Accelerator Laboratory, 2575 Sand Hill Road, Menlo Park, CA 94025, USA}

\author{M.F.~\surname{Kasim}}
\affiliation{Department of Physics, Clarendon Laboratory, University of Oxford, Parks Road, Oxford, OX1 3PU, UK}

\author{H.J.~\surname{Lee}}
\affiliation{Linac Coherent Light Source, SLAC National Accelerator Laboratory, 2575 Sand Hill Road, Menlo Park, CA 94025, USA}

\author{A.J.F.~\surname{Miscampbell}}
\affiliation{Department of Physics, Clarendon Laboratory, University of Oxford, Parks Road, Oxford, OX1 3PU, UK}

\author{B.~\surname{Nagler}}
\affiliation{Linac Coherent Light Source, SLAC National Accelerator Laboratory, 2575 Sand Hill Road, Menlo Park, CA 94025, USA}

\author{R.~\surname{Royle}}
\affiliation{Department of Physics, Clarendon Laboratory, University of Oxford, Parks Road, Oxford, OX1 3PU, UK}

\author{J.S.~\surname{Wark}}
\affiliation{Department of Physics, Clarendon Laboratory, University of Oxford, Parks Road, Oxford, OX1 3PU, UK}

\author{S.M.~\surname{Vinko}}
\email[]{sam.vinko@physics.ox.ac.uk}
\affiliation{Department of Physics, Clarendon Laboratory, University of Oxford, Parks Road, Oxford, OX1 3PU, UK}

\date{\today}

\begin{abstract}
The development of bright free-electron lasers (FEL) has revolutionised our ability to create and study matter in the high-energy-density (HED) regime. Current diagnostic techniques have been successful in yielding information on fundamental thermodynamic plasma properties, but provide only limited or indirect information on the detailed quantum structure of these systems, and on how it is affected by ionization dynamics. Here we show how the valence electronic structure of solid-density nickel, heated to temperatures of around 10 of eV on femtosecond timescales, can be probed by single-shot resonant inelastic x-ray scattering (RIXS) at the Linac Coherent Light Source FEL. The RIXS spectrum provides a wealth of information on the HED system that goes well beyond what can be extracted from x-ray absorption or emission spectroscopy alone, and is particularly well-suited to time-resolved studies of electronic-structure dynamics.
\end{abstract}

\maketitle

Understanding the quantum behaviour of high energy-density (HED) systems is of broad importance to applications across plasma physics~\cite{Ciricosta2012,Vinko2014}, astrophysics and planetary studies~\cite{Eggert:2010aa,Smith:2018aa}, and fusion energy science~\cite{Gaffney:2018,Vinko:2020}. This is largely due to the importance of particle correlations and quantum effects in governing ground- and excited-state plasmas at high density, including collisional plasma dynamics~\cite{Vinko2015,quincy:2018}, transport properties~\cite{Zylstra:2015,Desjarlais:2017,Ping:2019}, x-ray opacities~\cite{Bailey:2015aa,Preston:2017,Hollebon:2019} and the equation of state~\cite{Loubeyre:1996aa,Clerouin:2012}. Due to their large penetration depths, x-rays are often used to interrogate dense plasmas via spectroscopic techniques, for example via x-ray Thomson scattering (XRTS)~\cite{Glenzer:2009} to diagnose warm-dense plasmas~\cite{Garcia-Saiz:2008aa} and measure electron temperatures and densities~\cite{Kritcher69,Visco:2012}, ionization~\cite{Fletcher2014}, and ion temperatures and plasma viscosity~\cite{Mabey:2017aa,Larder:2019}. Hotter plasmas, with higher atomic numbers and high degrees of partial ionization, are in turn better suited to be interrogated via x-ray emission~\cite{Hoarty2013,Vinko:2015} or absorption spectroscopy~\cite{Recoules:2009,Bailey:2015aa,Ping:2016}. In contrast, resonant inelastic x-ray scattering (RIXS), while prevalent in atomic and condensed matter physics investigations~\cite{DellAngela:2013aa,Vaz-da-Cruz:2019aa,Couto:2017aa}, has not been used to date in HED research. This can be traced to three important requirements for RIXS: the x-ray source must be spectrally bright, spectrally stable, and have a tunable (narrow-bandwidth) wavelength to match the required resonance condition ~\cite{Gelmukhanov:1999,DellAngela:2016aa}. Historically, HED experiments have not had access to x-ray sources with such properties. The advent of x-ray free-electron lasers (FEL) now provides the capabilities to both create and probe extreme plasmas on ultra-short, inertially confined timescales~\cite{Vinko:2012aa,Fletcher:2015aa}, and, as we will show, allows for the application of RIXS-based techniques to the study of dense plasmas for the first time. In particular, we show how RIXS can be used to extract information on the electron temperature, ionization, valence band structure, and 1$s$ ionization energy (K-edge) of a warm-dense Ni plasma, all in a single, femtosecond, x-ray shot.

\begin{figure}
\includegraphics[width=1.0\linewidth]{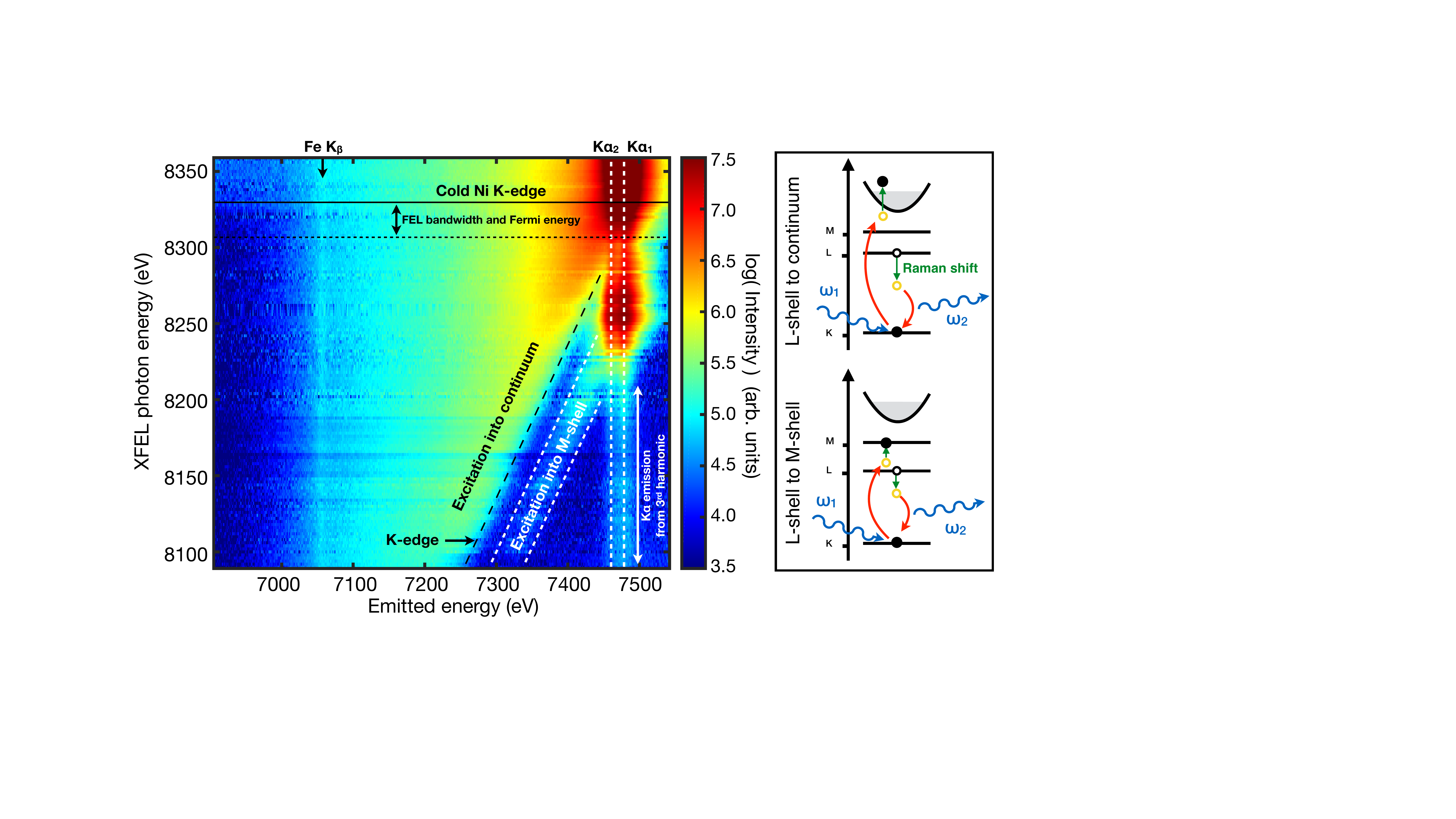}
\caption{Experimental RIXS map of the electronic valence structure of warm-dense Ni, irradiated at $1.2 \times 10^{17}$~Wcm$^{-2}$.
Spurious Fe K$_{\beta}$ emission at $\sim$7060~eV is due to small amounts of Fe impurities. The schematic shows the two RIXS channels, for excitations into M-shell states and into the continuum.}
\label{FIG:RIXS_annotated}
\end{figure}

\begin{figure*}
\includegraphics[width=\linewidth]{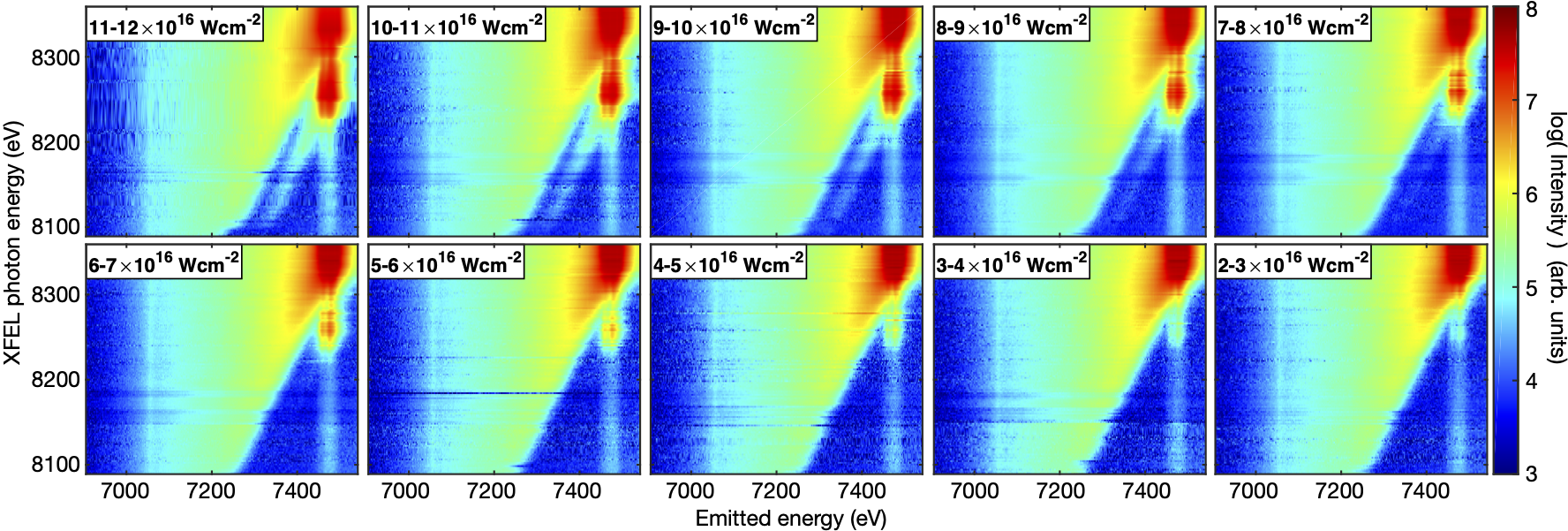}
\caption{RIXS map as a function of irradiation x-ray intensity. The x-ray pulse drives isochoric heating so higher intensities create hotter, more ionized plasma conditions. The intensity of the M-shell feature is strongly dependent on the x-ray intensity.}
\label{FIG:RIXS_intensity}
\end{figure*}

The experiment was carried out at the Matter in Extreme Conditions endstation~\cite{nagler2015matter} of the LCLS FEL. Here, x-rays at photon energies of 8.0 -- 8.4~keV were focused onto 25~$\mu$m thick Ni foils to spot diameters between 2-6~$\mu$m, producing x-ray intensities of $10^{16}-10^{17}$~Wcm$^{-2}$ on target. The thickness of the target was chosen to be one absorption length for irradiation below the Ni K-edge in order to maximize the total scattering signal while limiting the heating gradient in the longitudinal direction. The duration of the x-ray pulse was 60~fs FWHM, with a spectral bandwidth of 0.3\%.
The scattering signal was measured using a cylindrically curved highly annealed pyrolytic graphite (HAPG) crystal spectrometer in the von Hamos geometry, coupled to an Epix100 x-ray camera~\cite{carini2014studies}. The resulting spectral resolution of the instrument is 9~eV in our window of interest~\cite{nagler2015matter,zastrau2013characterization}.

We show the experimental scattering measurements obtained at the highest irradiation intensity in Fig.~\ref{FIG:RIXS_annotated}. The LCLS photon energy, shown on the ordinate axis, is tuned below and across the K-edge of Ni, illustrating a rich spectrum of down-scattered photons mapping out the valence electronic structure. The dominant RIXS channels are shown schematically in the adjacent diagram. Given the short-lived $1s$ core-hole state, and high dynamic range of the spectrometer, we find our setup is sensitive to scattering over a spectral window of nearly 500~eV, capturing features across a wide band of valence states. When the x-ray pulse is tuned to be below both the K-edge and the K-M resonances ($\sim$8250~eV), no absorption can take place in the K-shell (1$s$ states), resulting in no K$_{\alpha}$ radiative $2p-1s$ recombination. While absorption and emission processes are thus suppressed, inelastic scattering can efficiently occur in the vicinity of absorption resonances, resulting in photons down-scattered in energy. We identify two main channels for this process: both lead to an L-shell vacancy in the final RIXS state, but one excites an electron into the M-shell ($3p$ state), while the other excites an electron into the continuum.
We note that experimentally we always observe some weak K$_{\alpha}$ emission, due to the presence of the unfocused third harmonic of the XFEL ($>$24~keV, containing$\sim 0.1\%$ of the total FEL energy). However, this contribution is readily distinguished from that of the fundamental wavelength as it only contributes a weak, source-broadened background to the K$_{\alpha}$ feature.

The RIXS cross section for scattering into a solid angle $d\Omega$ can be written as~\cite{kramers1925streuung,kotani2001resonant}:
\begin{align}
\frac{d^2\sigma}{d\Omega d\omega_2} \propto \sum_f \left |  \sum_n \frac{\langle f | \mathcal{D^{\prime \dagger}} | n \rangle   \langle n | \mathcal{D} | i \rangle }{E_n - \hbar \omega_1 - E_i + i\Gamma} \right |^2 \delta(E_f - E_i + \hbar \tilde \omega),
\end{align}
where $|i\rangle$, $|n\rangle$ and $|f\rangle$ are initial, intermediate (with a core-hole), and final states of the system, with respective energies $E_i$, $E_n$ and $E_f$. The incident and outgoing photons have energies $\hbar \omega_1$ and $\hbar \omega_2$, and $\tilde \omega = \omega_2 - \omega_1$.
The transition operator $\mathcal{D}$ is taken in the dipole approximation $\mathcal{D}\approx \mathbf{p} \cdot \boldsymbol{\eta}$, with $\mathbf{p}$ the electron momentum and $\boldsymbol{\eta}$ the photon polarization vector.
Much of the complexity in calculating the RIXS cross section arises from the coupling of the core hole potential with the intermediate states, and in determining how this excited system evolves before recombination. However, here we are not interested in following these low energy excitations, which are limited by the inverse of the core-hole lifetime, equivalent to a few eV for Ni. Instead, we seek to obtain a broad picture of the unoccupied valence states, and demonstrate that the overall electronic structure can be extracted efficiently from the measurement. Neglecting core-hole interactions, our RIXS cross section can thus be simplified significantly to (see supplementary material)
\begin{align}
\frac{d^2\sigma}{d\Omega d\omega_2} \propto  \sum_f
\frac{[1-f_{\rm FD}(\varepsilon;T)] \rho (\varepsilon) \; A_f \left | M(\varepsilon)\right|^2} {(\hbar \omega_2-(\epsilon_{L,f}-\epsilon_K))^2+\Gamma_f^2}.
\label{Eq:XS2}
\end{align}
Here $\varepsilon = \hbar\omega_1-\hbar\omega_2+\epsilon_{L,f}$, with $\epsilon_{L,f}$ the energy of the final L-shell core hole, and the sum is over all final L-shell hole states.
$A_f$ represents the K$_{\alpha}$ line intensity, $\Gamma_f$ the linewidth, $M(\varepsilon)$ the dipole matrix elements for the transition of a $1s$ electron into a valence state, $\rho (\varepsilon)$ the final density of states (DOS), and $f_{\rm FD}(\varepsilon;T)$ the Fermi distribution at temperature $T$.
We see that if we can neglect core-hole interactions on the intermediate state, the RIXS spectrum can be approximately described as an absorption cross section ($\propto \rho (\varepsilon) | M(\varepsilon)|^2$) modulated by the Lorentzian width of the K$_{\alpha}$ recombination line. To compare with experiment we have evaluated the transition matrix elements and the DOS using finite-temperature density functional theory (DFT)~\cite{Perdew:1996p3865,Monkhorst:1976p5188} via the codes ABINIT~\cite{torrent2008implementation,Gonze2009,Gonze2016a} and ATOMPAW~\cite{Holzwarth:2001p6935}. Electron binding energies, line ratios, and lifetimes were taken from the Henke tables~\cite{henke1993x}. 

Making use of Eq.~\ref{Eq:XS2}, we see that in principle the spectrum provides access to the electron temperature via the shape of the Fermi distribution, to the density of vacant states in the valence band via the scattering intensity, and to the K-edge energy. By comparing the latter with known atomic binding energies the ionization potential depression can thus be extracted from the data as a function of plasma conditions.

For RIXS to take place, a vacant final state is needed for the electron excitation~\cite{ament:2011}. In the ground state the only available states for exited electrons are above the Fermi energy, 8333~eV above the $1s$ state. Excitation between fully occupied inner-shell states cannot occur. However, if the system is heated and partially ionized so that vacancies appear in some bound states, additional inner-shell excitation channels open up, leading to new features in the RIXS spectrum. We show this effect in Fig.~\ref{FIG:RIXS_intensity} where we present the same RIXS map as in Fig.~\ref{FIG:RIXS_annotated}, but for 10 different x-ray intensities spanning nearly an order of magnitude. By changing the intensity on target we can tune the degree of isochoric heating and inner-shell ionization, producing different systems diagnosed by RIXS. The main K$_{\alpha}$ emission features are present in all spectral maps, and are due to above-edge photoionization and photoexcitation. In contrast, the intensity of the inelastic scattering signal resulting in a M-shell excitation shows a strong dependence on the XFEL intensity. At irradiation intensities $\lesssim 7 \times 10^{16}$~Wcm$^{-2}$ there is little observable scattering from transitions into M-shell states, although for intensities down to $4 \times 10^{16}$~Wcm$^{-2}$ the M-shell resonance can still be seen as an enhanced K$_{\alpha}$ feature at 8250~eV. This corresponds to a process whereby the XFEL photon resonantly drives a $1s-3p$ transition, followed by radiative $2p-1s$ K$_{\alpha}$ recombination. At intensities $\gtrsim 7 \times 10^{16}$~Wcm$^{-2}$ however, and for photon energies below the resonant $1s-3p$ transition energy, we observe a strong new inelastic feature corresponding to a photon down-scattered due to an inner-shell L-M excitation. The intensity of this feature corresponds to the M-shell states changing from being purely virtual to real, due to ionization. This signal thus not only shows the position of the M-shell atomic states, but also provides a highly sensitive measurement of bound-state thermal ionization.
The experimental spectra in Figs.~\ref{FIG:RIXS_annotated} and \ref{FIG:RIXS_intensity} were averaged according to the central XFEL frequency in bins of 2~eV in width and normalized to incident pulse energy, with a total of 11,383 x-ray shots included. This results in between one and 63 shots per bin, populated by tuning the XFEL photon energy in 20~eV increments with a $\pm$10~eV stochastic spread.
This ability to collect single-shot data is essential for HED studies where significant shot-to-shot variations can occur. This is particularly relevant in the context of time-resolved studies of matter driven to high temperatures or pressures via optical lasers, where uncertainties in the generated conditions limit the information that can be extracted from integrated experiments~\cite{Harmand:2015}.

\begin{figure}
\includegraphics[width=1.0\linewidth]{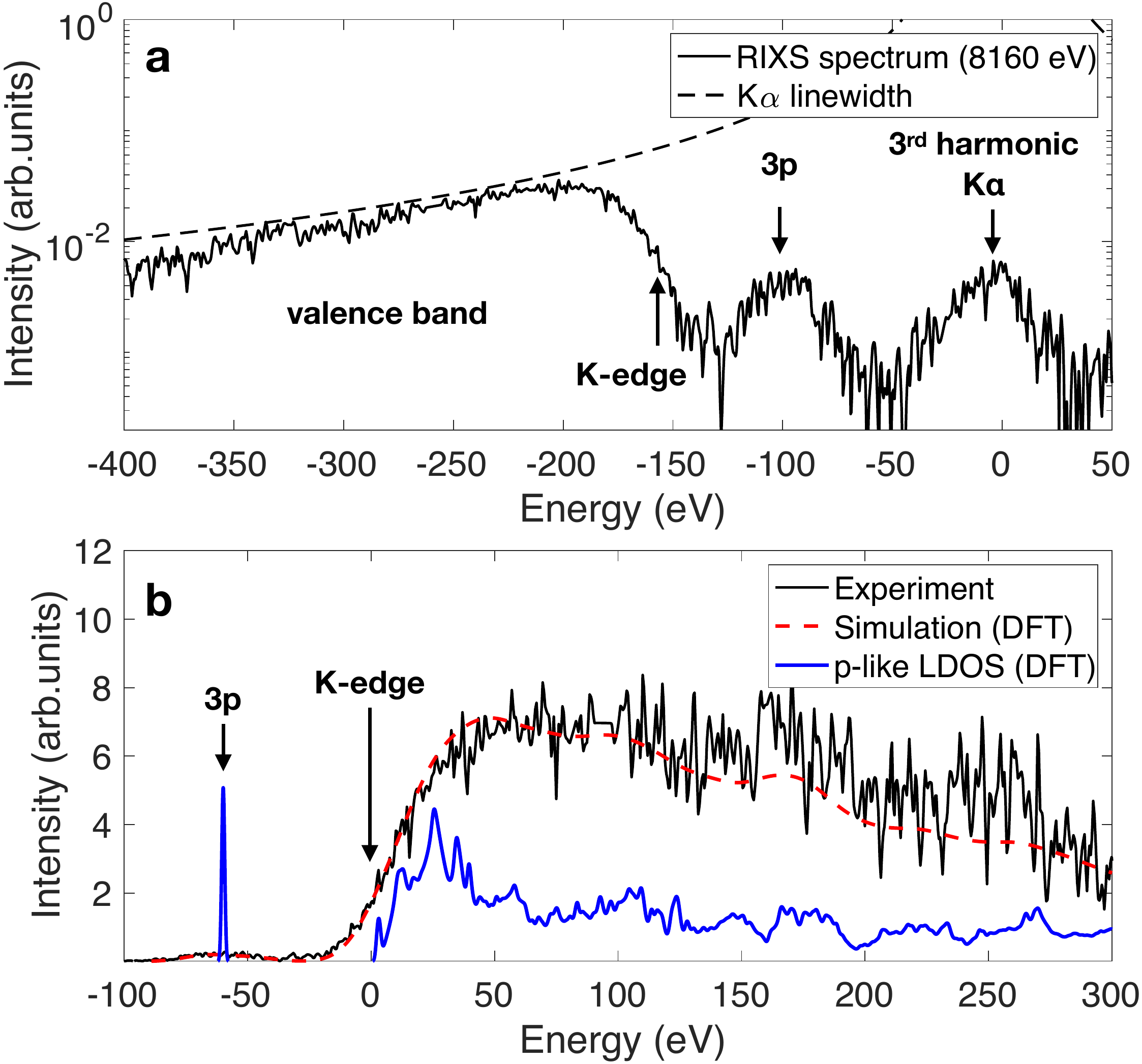}
\caption{a) RIXS spectrum measured from the K$_{\alpha}$ resonance (FEL photon energy of 8160~eV, peak intensity $1.2 \times 10^{17}$~Wcm$^{-2}$). Features from the K-edge and vacant 3$p$ states are highlighted, alongside the envelope of the K$_{\alpha}$ emission, for reference. b) Absorption part of the RIXS spectrum after removing the K$_{\alpha}$ emission contribution (see Eq.~\ref{Eq:XS2}), compared with DFT simulations. Also shown is the Ni $p$-like LDOS before experimental broadening, illustrating the contributing RIXS states. Zero energy is set to the bottom of the valence band.}
\label{FIG:Lineouts}
\end{figure}

We show a lineout of the RIXS spectrum for a system driven at 8160~eV and at the highest intensity in Fig.~\ref{FIG:Lineouts}a. The spectrum shows the RIXS features due to excitations into $p$-like valence states, and the 3rd harmonic contribution to the K$_{\alpha}$ emission line.
By removing the Lorentzian core-hole lifetime contribution to the spectrum we extract the absorption component, and compare it with our DFT calculations in Fig.~\ref{FIG:Lineouts}b. We find good agreement between experiment and simulation, but note the comparison is limited by the spectral resolution of the data, a combination of spectrometer resolution (9~eV) and FEL bandwidth ($\sim$20~eV). The x-ray absorption cross section is closely related to the angular momentum projected density of states (LDOS), the two being proportional to within a smoothly varying function of the energy~\cite{Muller:1984,Muller:1998}. As such, the measurement (at low energies) can be well approximated simply by the $p$-like LDOS, which we plot (unbroadened) alongside the full transition matrix elements calculation in Fig.~\ref{FIG:Lineouts}b. The experimental resolution is sufficient to accurately extract the area under the $3p$ feature to measure ionization, and to determine the position of the K-edge to within a few eV, but does not allow for an accurate measurement of the temperature directly via the Fermi edge. However, this is purely a limitation of our setup, optimized to maximise the intensity of the RIXS spectrum at the expense of the resolution.

\begin{figure}
\includegraphics[width=1.0\linewidth]{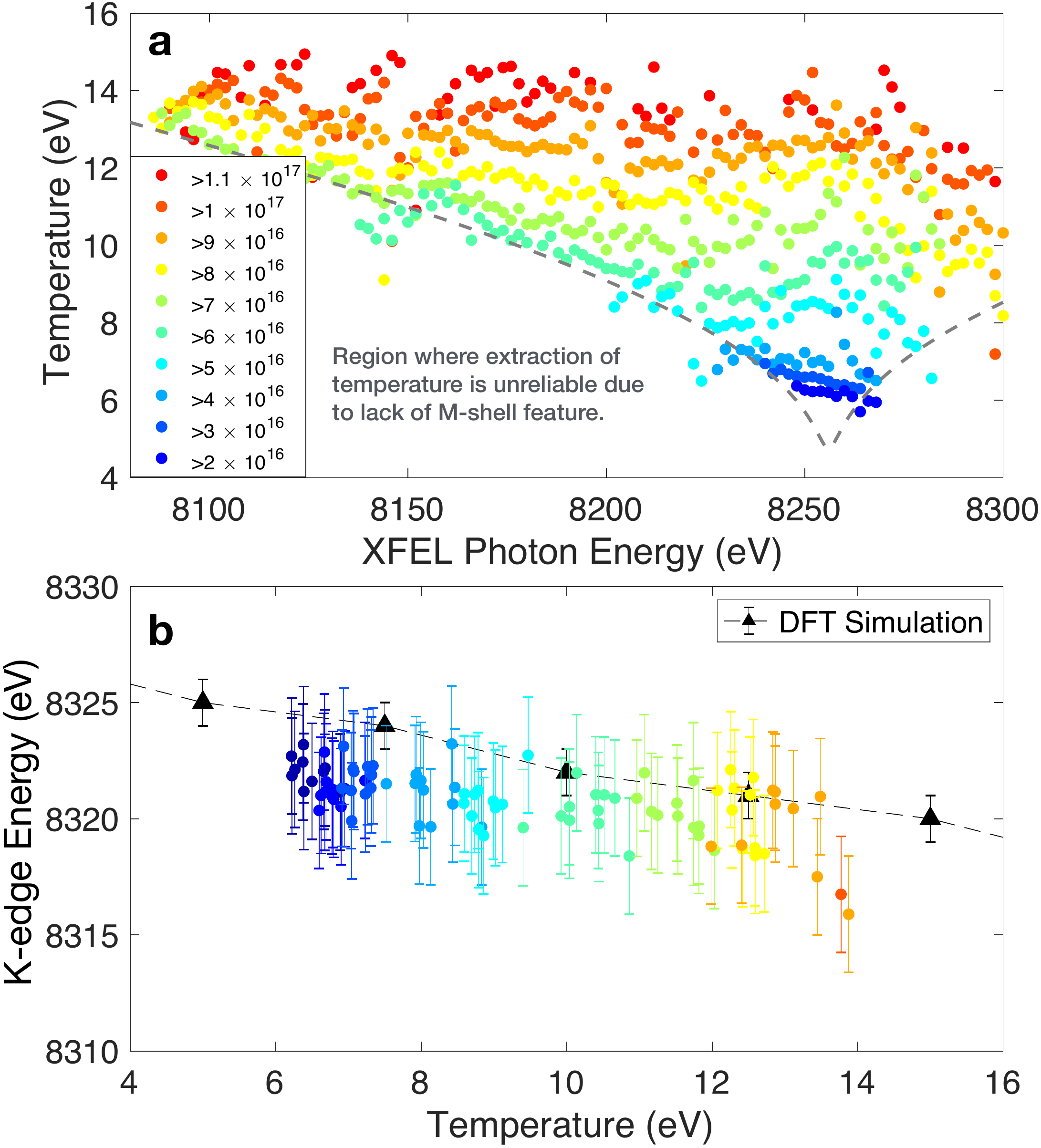}
\caption{(a) Temperature and (b) K-edge position extracted from the RIXS dataset at different irradiation intensities. K-edge energies are compared with finite-temperature DFT calculations. Temperature uncertainties are $\pm$1-2 eV, not plotted for clarity.}
\label{FIG:IPD}
\end{figure}                                                                                                                                                                                                                                                                                                                                                                                                                                                                                                                                                                 

By fitting the cross section of Eq.~\ref{Eq:XS2} to the full experimental dataset we can proceed to extract physical parameters for the transient Ni system as a function of x-ray isochoric heating. We show these results in Fig.~\ref{FIG:IPD}. We assume that the valence electrons are close to thermal equilibrium~\cite{delaVarga:2013,quincy:2018}, so the temperature can be deduced directly from the degree of M-shell ionization. Temperature uncertainties extracted in this manner are between $\pm$1-2 eV, and are seen to change little with photon energy. 
However, they depend strongly on the intensity of x-ray pulse, which is consistent with higher final energy-densities reached by higher intensity x-ray pulses: peak temperatures between 6 and 15~eV are found for irradiation intensities between $2 \times 10^{16}$ and $\sim 10^{17}$~Wcm$^{-2}$. The temperature values extracted are a good indication of the peak temperature in the target, rather than of the average, as the ionization has a near exponential dependence on temperature while $T\ll E_{bind}$, so the peak ionization will strongly dominate the emission. We note that the RIXS signal is extremely sensitive to ionization, and clear M-shell features appear in the data for $3p$ ionization fractions well below 1\%. From Fig.~\ref{FIG:IPD}b we can observe a slight variation of the position of the K-edge with ionization and temperature. As expected, the K-edge decreases with temperature, and its overall position agrees well with our DFT simulations.

The RIXS process couples an x-ray absorption and emission event. The spectrum corresponding to an inner-shell electron excitation thus contains the same information on the unoccupied LDOS as an x-ray absorption spectroscopy measurement~\cite{Nagasawa:1989}, but, depending on experimental setup, with several potential advantages in terms of symmetry projection selectivity, access to higher multipole excitations, and the absence of lifetime broadening~\cite{Schuelke:RIXS}. Importantly, RIXS enables a broadband absorption spectroscopy measurement to be made using a very narrow bandwidth probe. In fact, the range over which the absorption properties of the vacant states can be measured do not depend on the bandwidth of the x-ray source at all, only on the lifetime of the intermediate state. For a system with very short-lived core-holes, as is the case for high-Z ions, broad absorption spectra can be collected. Perhaps counterintuitively, the bandwidth of the source only acts to limit the resolution of the absorption features that can be observed, and should thus be made as small as possible. This observation is promising because the narrow bandwidths of typical FEL sources have so far placed stringent limitations on absorption spectroscopy diagnostics in high energy density experiments, in particular where single-shot information is required~\cite{Principi:2016}. Our results show that the peak spectral brightness of an FEL pulse is sufficiently large to enable a single-shot vacant LDOS investigation via RIXS instead.

In summary, using RIXS, we have demonstrated the high potential of a novel approach to investigate the valence electronic structure of warm-dense transient systems. Our methods let us access core electron ionization and valence electronic structure simultaneously, in a femtosecond snapshot. We have found the K-edge energy of solid-density Ni heated to 10-15 eV temperatures, and have tracked its shifts for small M-shell ionizations -- less than 1\%. This is an important addition for directly probing electronic-structure dynamics of extreme states of matter, and the single-shot nature of the technique lends itself well to future time-resolved studies.

\begin{acknowledgments}
Use of the Linac Coherent Light Source (LCLS), SLAC National Accelerator Laboratory, is supported by the U.S. Department of Energy, Office of Science, Office of Basic Energy Sciences under Contract No. DE-AC02-76SF00515.
The MEC instrument was supported by the U.S. Department of Energy, Office of Science, Office of Fusion Energy Sciences under contract no. DE-AC02-76SF00515.
O.S.H. acknowledges support from the Oxford University Centre for High Energy Density Physics (OxCHEDS).
R.S.M. acknowledges support under the NSERC Discovery Grant program.
O.S.H., Q.vdB and S.M.V. acknowledge support from the Royal Society.
M.F.K, R.R., J.S.W. and S.M.V. acknowledge support from the U.K. EPSRC under grant EP/P015794/1.
S.M.V. is a Royal Society University Research Fellow.
\end{acknowledgments}


%\bibliographystyle{apsrev4-1}
%\bibliography{Refs}

\end{document}